\documentclass[12pt]{iopart}

\usepackage{iopams}
\usepackage{graphicx}
\begin{document}

\title{Spin interactions in a quantum dot containing a magnetic impurity}

\author{Aram Manaselyan and Tapash Chakraborty}

\address{Department of Physics and Astronomy, University of
Manitoba, Winnipeg, Canada R3T 2N2}
\ead{tapash@physics.umanitoba.ca}

\begin{abstract}
The electron and hole states in a CdTe quantum dot containing a
single magnetic impurity in an external magnetic field are
investigated, using the multiband approximation which includes the
heavy hole-light hole coupling effects. The electron-hole spin
interactions and s,p-d interactions between the electron, the hole
and the magnetic impurity are also included. The exciton energy
levels and optical transitions are evaluated using the exact
diagonalization scheme. A novel mechanism is proposed here
to manipulate impurity-spin in the quantum dot which allows us to
drive selectively the spin of the magnetic atom into each of its
six possible orientations.

\end{abstract}

\submitto{\NT}
\maketitle

\section{Introduction}
Quantum dots (QDs), the nanoscale zero-dimensional systems with
discrete energy levels, much like in atoms (and hence the popular
name, artificial atoms \cite{qdbook}) have one great advantage that
the shape and the number of electrons and holes in those systems can
be controlled externally and as a result, these systems have been
the subject of intense research in recent years. In the QDs it is
therefore possible to localize single spin in an area of a few
nanometers. Conservation of the angular momentum in the QD allows
for both spin sensitive detection and injection. By coupling these
techniques with ultrafast optical pulses, it is possible to
stroboscopically measure the spin dynamics in a QD. Thus QDs are
particularly promising as components of futuristic devices for
quantum information processing and for coherent spin transport
\cite{SAQD}.

Detection and control of single magnetic atoms in QDs represent the
fundamental scaling limit for the magnetic information storage. The
strong and electronically controllable spin-spin interactions that
exists in magnetic semiconductors offer an ideal laboratory for
exploring the single magnetic spin readout and control. Magnetic
ions placed within a lattice can exhibit relatively long spin
lifetimes \cite{Spintronics}. Exploration of single magnetic spins
in semiconductor QDs was pioneered in (II,Mn)-VI systems where the
magnetic atoms are isoelectronic Mn$^{2+}$ ions with spin 5/2 (see
e.g. \cite{Bhattacharjee2003,Govorov2004,Peeters2008}). A different
situation arises in (III,Mn)-V magnetic semiconductors, where the
Mn$^{2+}$ ions contribute to the acceptor states within the band gap
causing the magnetic ions to behave as optical spin centers
\cite{Bhattacharjee2008}. Experiments have clearly illustrated the
effect of magnetic ions on exciton optical spectrum in the QDs
\cite{Besombes2005}. Theoretically, the effect of spin-exciton
interactions on the optical spectrum of a quantum dot with a
magnetic impurity is considered in \cite{Rossier2006}. But due to
approximations used in \cite{Rossier2006}, the results are limited
only to the case of zero or low magnetic fields. Recently Gall et.
al. \cite{Besombes2009} presented a new way to optically probe the
spin of a single magnetic impurity in the QD. They demonstrated that
optical excitation of an individual Mn-doped QD with circularly
polarized photons can be used to prepare nonequilibrium distribution
of the Mn spin, even in the absence of an applied magnetic field.
Reiter et. al. \cite{Reiter} have presented a technique for an
all-optical switching of the spin state of a magnetic atom in a QD
on a picosececond time scale. They have shown that the spin state of
a single Mn atom in a QD can be selectively controlled by
manipulating the exciton states with ultrafast laser pulses. All six
possible spin states can be reached. The switching process can be
optimized by applying a magnetic field.

In this paper we report on our theoretical studies involving
electron and hole states in a CdTe quantum dot containing a single
magnetic impurity in an external magnetic field. Here we show that
the s,p-d spin interaction brings about level anticrossings between
the dark and bright exciton states. We explain the physics behind
these anticrossings. Our results are in good agreement with the
experimental observations \cite{Besombes2005}. We also propose a new
magneto-optical mechanism for manipulation of the magnetic impurity
spin, by using the laser pulses and by varying the strength of the
magnetic field.

\section{Theory}
In our investigation of the electron and hole states in a
cylindrical CdTe quantum dot with a single magnetic impurity,
subjected to a perpendicular magnetic field, we choose the lateral
confinement potential of the dot as parabolic with the corresponding
frequencies $\omega_0^e$ and $\omega_0^h$ for electron and hole
respectively. This choice can be justified from the energies of the
far-infrared absorption on such dots, which shows only a weak
dependence on the electron occupation \cite{Parabol,Warburton}. We
also take into account the confinement potential in the growth
direction as a rectangular well of width $L$. We assume that the
size of the dot is smaller than the bulk exciton Bohr radius and neglect
the long-range electron-hole Coulomb interaction \cite{Efros,
Bhattacharjee2007}. The Hamiltonian of the system can then be
written as
\begin{equation}\label{Ham}
{\cal H}={\cal H}_e+{\cal H}_h+{\cal H}_{s-d}+{\cal H}_{p-d}+{\cal
H}_{eh},
\end{equation}
where ${\cal H}_{s-d}=-J_e \delta({\bf r}_e-{\bf
r}_{Mn})\boldsymbol{\sigma} {\bf S}$ and ${\cal H}_{p-d}=-J_h
\delta({\bf r}_h-{\bf r}_{Mn}){\bf j} {\bf S}$ describe the electron-Mn
and hole-Mn spin-spin exchange interaction with strengths $J_e$ and
$J_h$ respectively. ${\cal H}_{eh}=-J_{eh}\delta({\bf r}_e-{\bf
r}_h)\boldsymbol{\sigma} \bf{j}$ is the electron-hole spin
interaction Hamiltonian \cite{Efros}.

The electron Hamiltonian is
\begin{equation}\label{eHam}
{\cal H}_e=-{\frac 1 {2m_e}} \left( {\bf p}-{\frac e c} {\bf
A}\right)^2+V_{conf}^e (\rho, z)+ \frac12 g_e \mu_B B \sigma_z,
\end{equation}
where ${\bf A}=B/2(-y,x,0)$ is the vector potential of magnetic
field in the symmetric gauge and the last term is the electron Zeeman
energy. The eigenfunctions of ${\cal H}_e$ can be written as
\begin{equation}\label{eBazis}
\psi_{nls\sigma}^e(\rho,\theta,z)=f_{nl}^e(\rho)e^{il\theta}
g_s(z)\chi_\sigma,
\end{equation}
where $\chi_\sigma$ are the electron spin functions,
$$g_s(z)=\sqrt{\frac 2 L}\sin\left[{\frac {s\pi} L}\left(z+
{\frac L 2}\right)\right] \qquad
s=1,2,3,\ldots$$ and the in plane functions are Fock-Darwin orbitals
\cite{qdbook}
\begin{equation}
f_{nl}(\rho)=C_{nl}e^{-\frac {\rho^2} {2a_e^2}}\left(\frac{i\rho}
{a_e}\right)^{|l|}L_n^{|l|}\left(\frac{\rho^2}
{a_e^2}\right)\end{equation}
\begin{displaymath}
C_{nl}=\frac1a\left[\frac{n!}{\pi(n+|l|)!}\right]^{\frac12}, \quad
n=0,1,2,\ldots, \quad l=0,\pm1,\pm2,\ldots
\end{displaymath}

The single-electron energy is given by
\begin{displaymath}
E_{nls\sigma}=2\hbar \omega_e \left(n-l \frac{\omega_c^e}{4\omega_e}+
\frac{|l|+1}{2}\right)+\frac{\hbar^2\pi^2s^2}{2m_eL^2}+
g_e\mu_B B\sigma,
\end{displaymath}
where $\omega_c^e=eB/m_ec$ is the cyclotron frequency,
$\omega_e=\sqrt{(\omega_o^e)^2+0.25(\omega_c^e)^2}$ and
$a_e=\sqrt{\hbar/m_e\omega_e}$.

Taking into account only the $\Gamma_8$ states which correspond to
the states with the hole spin $j=3/2$ and include the heavy
hole-light hole coupling effects, we can construct the single-hole
Hamiltonian in the dot as
\begin{equation}\label{hHam}
{\cal H}_h={\cal H}_L+V_{conf}^h(\rho,z)-2\kappa \mu_B B j_z.
\end{equation}
Here ${\cal H}_L$ is the Luttinger hamiltonian in axial
representation obtained with the four-band \textbf{k$\cdot$p} theory
\cite{Lutt2,Pedersen2}
\begin{equation}\label{HLut}
{\cal H}_L=\frac{1}{2m_0}\left( \begin{array}{cccc}
{\cal H}_h & R & S & 0 \\
R^* & {\cal H}_l & 0 & S \\
S^* & 0 & {\cal H}_l & -R \\
0 & S^* & -R^* & {\cal H}_h
\end{array}\right),
\end{equation}
where
\begin{eqnarray*}
{\cal H}_h & = & (\gamma_1+\gamma_2)(\Pi_x^2+\Pi_y^2)+
(\gamma_1-2\gamma_2)\Pi_z^2,\\
{\cal H}_l & = & (\gamma_1-\gamma_2)(\Pi_x^2+\Pi_y^2)+
(\gamma_1+2\gamma_2)\Pi_z^2,
\end{eqnarray*}
$R=2\sqrt3\gamma_3{\rm i}\Pi_-\Pi_z, \quad S=\sqrt3\gamma \Pi_-^2,
\quad \gamma=\frac12(\gamma_2+\gamma_3),$ and ${\bf \Pi}={\bf
p}-\frac{e}{c}{\bf A}, \quad \Pi_\pm=\Pi_x\pm{\rm i}\Pi_y.$
$\gamma_1, \gamma_2$, $\gamma_3$ and $\kappa$ are the Luttinger
parameters and $m_0$ is the free electron mass.

The Hamiltonian (\ref{hHam}) is rotationally invariant. Therefore it
will be useful to introduce the total momentum ${\bf F}={\bf j}+{\bf
l}$, where ${\bf j}$ is the angular momentum of the band edge Bloch
function, and ${\bf l}$ is the envelop angular momentum. Since the
projection of the total momentum $F_z$ is a constant of motion, we
can find simultaneous eigenstates for (\ref{hHam}) and $F_z$
\cite{Sersel}.

For the given value of $F_z$ it is logical to seek the
eigenfunctions of Hamiltonian (\ref{hHam}) as an expansion
\cite{Pedersen2, Manaselyan}
\begin{equation}\label{hBazis}
\Psi_{F_z}(\rho,\theta,z)=\sum_{n,s,j_z}C_{F_z}(n,s,j_z)
f_{n,F_z-j_z}^h(\rho)e^{i(F_z-j_z)\theta}g_s(z)\chi_{j_z},
\end{equation}
where $\chi_{j_z}$ are hole spin functions and $f_{nl}^h(\rho)$ are
the Fock-Darwin orbitals for hole with
$a_h=\sqrt{\hbar(\gamma_1+\gamma_2)/m_0\omega_h}$,
$\omega_h=\sqrt{(\omega_o^h)^2+0.25(\omega_c^h)^2}$ and
$\omega_c^h=eB(\gamma_1+\gamma_2)/m_0c$. The matrix elements of the
Hamiltonian (\ref{hHam}) can then be evaluated analytically. All
single hole energy levels and expansion coefficients are evaluated
numerically using the exact diagonalization scheme
\cite{Manaselyan}. Calculations are carried out for the CdTe quantum
dot with sizes $a_e=a_h=37$ \AA, $L=25$\AA\ and with following
parameters: $E_g=1.59$ eV, $m_e=0.096 m_0$, $g_e=-1.5$
$\gamma_1=5.29$, $\gamma_2=1.8$, $\gamma_3=2.46$, $\kappa=0.7$
\cite{Adachi}.

To include spin-spin interactions, we can construct the wave
function of the electron, hole and the magnetic impurity as an
expansion of direct products of the lowest state wave functions
(\ref{eBazis}), (\ref{hBazis}) and eigenfunctions for the magnetic
impurity.
\begin{equation}\label{Bazis}
\Psi_{J_z}=\sum_\sigma \sum_{F_z} \sum_{S_z}
C(\sigma,F_z,S_z)\psi_{0,0,1,\sigma}^e\times \psi_{F_z}^h\times
|S_z\rangle.
\end{equation}
Here $\sigma=\pm1/2$, $S_z=\pm1/2, \pm3/2, \pm5/2$ and $F_z=\pm1/2,
\pm3/2,\pm5/2 \ldots$ and $J_z$ is the projection of the total
momentum ${\bf J}={\bf \sigma}+{\bf F}+{\bf S}$. Using the
components of this expansion as the new basis functions, we can
calculate the corresponding matrix elements for the electron-hole,
the electron-impurity and the hole-impurity interactions. For the
electron-hole interaction we have
\begin{eqnarray}\label{Meh}
M_{eh}=-2\pi J_{eh}
\delta_{S_z,S_z'}\sum_{nsj_z}\sum_{n's'j_z'}C_{F_z}(n,s,j_z)
C_{F_z'}(n',s',j_z')\delta_{F_z-j_z,F_z'-j_z'}\times
\nonumber\\
\int_0^\infty\left|f_{00}^e(\rho)\right|^2\left(f_{n,F_z-j_z}^h(\rho)
\right)^*f_{n',F_z'-j_z'}^h(\rho)\rho
d\rho \times \nonumber\\
\int_{-L/2}^{L/2}\left(g_1^e(z)\right)^2g_s^h(z)g_{s'}^h(z)dz \times
\left\langle\sigma,j_z|\boldsymbol{\sigma} {\bf
j}|\sigma',j_z'\right\rangle,
\end{eqnarray}
where $\boldsymbol{\sigma}$ is the Pauli spin operator and $\bf{j}$
is the hole spin operator with following components
\begin{eqnarray*}
j_x=\left(\begin{array}{cccc} 0 & \frac{{\rm i}\sqrt{3}}2
& 0 & 0\\
-\frac{{\rm i}\sqrt{3}}2 & 0 & {\rm i} & 0\\ 0 & -{\rm i} & 0
& \frac{{\rm i}\sqrt{3}}2\\
0 & 0 & -\frac{{\rm i}\sqrt{3}}2 & 0 \\
\end{array}\right), \qquad
j_y=\left(\begin{array}{cccc} 0 & \frac{\sqrt{3}}2 &
0 & 0\\
\frac{\sqrt{3}}2 & 0 & 1 & 0\\ 0 & 1 & 0 & \frac{\sqrt{3}}2\\ 0 &
0 & \frac{\sqrt{3}}2 & 0 \\
\end{array}\right), \\
j_z=\left(\begin{array}{cccc} \frac32 & 0 &
0 & 0\\
0 & \frac12 & 0 & 0\\ 0 & 0 & -\frac12 & 0\\ 0 &
0 & 0 & -\frac32 \\
\end{array}\right).
\end{eqnarray*}

For the case of electron-impurity interaction we get
\begin{equation}\label{Msd}
M_{s-d}=-\frac{J_e}{4\pi}\delta_{j_z,j_z'}\left|f_{00}^e(\rho_i)
\right|^2(g_1^e(z_i))^2\left\langle\sigma_z,S_z|
\boldsymbol{\sigma}{\bf{S}}|\sigma_z',S_z'
\right\rangle,
\end{equation}
where $\rho_i$ and $z_i$ are the impurity coordinates. Finally
for the case of hole-impurity interaction we get
\begin{eqnarray}\label{Mpd}
M_{p-d}=-\frac{J_h}{2\pi}\delta_{\sigma,\sigma'}\sum_{n,s,j_z}
\sum_{n's'j_z'}C_{F_z}(n,s,j_z)C_{F_z'}(n',s',j_z')
\times \nonumber\\
\left(f_{n,F_z-j_z}^h(\rho_i)\right)^*f_{n',F_z'-j_z'}^h(\rho_i)
g_s^h(z_i)g_{s'}^h(z_i)
\left\langle j_z,S_z|{\bf{j}}{\bf{S}}|j_z',S_z' \right\rangle.
\end{eqnarray}
In order to calculate spin matrix elements in (\ref{Msd}, \ref{Mpd})
we need to introduce the raising and lowering operators $\bf{S_+}$
and $\bf{S_-}$.
\begin{eqnarray}
{\bf{S_+}}|S_z\rangle=\sqrt{S(S+1)-S_z(S_z+1)}|S_z+1\rangle,
\nonumber\\
{\bf{S_-}}|S_z\rangle=\sqrt{S(S+1)-S_z(S_z-1)}|S_z-1\rangle.
\end{eqnarray}

When the magnetic impurity is located at
the center of the dot, we have non vanishing matrix elements for the
hole-impurity interaction only for hole states with $F_z=\pm3/2$ and
$\pm1/2$. Therefore the number of basis states for that case is 48.
Since the hole ground state and the first few low-lying states are
described by $F_z=\pm3/2$ and $\pm1/2$, we can use the same 48 basis
states also for the case of off-center impurity. The problem was
solved numerically using the exact diagonalization scheme and with
interaction parameters $J_e=15$ meV nm$^3$, $J_h=-60$ meV nm$^3$
\cite{Besombes2005, Reiter}.

\section{Discussion}
In the absence of a magnetic atom and without the electron-hole spin
interaction, the hole ground state is at $F_z=\pm3/2$ and the ground
state of the electron-hole pair will be four-fold degenerate with
values of total momentum $J_z=\pm1$ and $\pm2$. The magnetic field
lifts that degeneracy due to the Zeeman splitting and as a result
two bright ($J_z=\pm1$) and two dark ($J_z=\pm2$) exciton states
appear. The electron-hole exchange interaction in turn gives rise to
a further splitting between the bright and dark exciton states and
removes the degeneracy between them at zero magnetic field.

\begin{figure}
\includegraphics{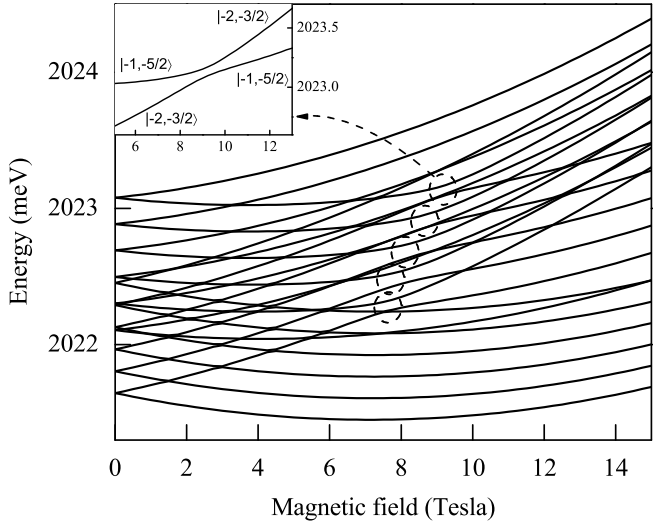}
\caption{\label{fig:Energy} Magnetic field dependence of the
electron and hole total energy levels taking into account the sp-d
interaction with a magnetic impurity located at the center of the
dot. The detailed picture of one of the anticrossings due to spin
interaction is presented as inset.}
\end{figure}

The sp-d exchange interaction between the electron (hole) and the
magnetic impurity will split each of these four exciton energy
levels to six. As a result, there are 24 separate energy levels. In
figure~\ref{fig:Energy}, we show the electron-hole energy levels as
a function of the magnetic field with the magnetic impurity at the
center of the dot. Each state can be presented as a linear
superposition of the 48 basis functions defined above. For a given
value of the magnetic field, each state can be labeled by the most
important component of the basis states. As an example, for the
ground state the most important component is $\sigma=1/2$, $F_z=3/2$
and $S_z=-5/2$ for all values of the magnetic field. So we can label
it $|1/2,3/2,-5/2\rangle$ or $|2,-5/2\rangle$. With an increase of
the magnetic field we see many level crossings and anticrossings.
The most interesting ones are the five anticrossing points, marked
by circles in the figure. The reason for these anticrossings is the
sp-d interaction between the dark and bright states with same total
momentum. For example, the bright exciton state $|-1,-5/2\rangle$
will couple to the dark state $|-2,-3/2\rangle$. As a result, there
will be an anticrossing at a magnetic field of 9 Tesla, which is
presented as inset of figure~\ref{fig:Energy}. The other four
anticrossings are due to coupling of the states $|-1,-3/2\rangle$
and $|-2,-1/2\rangle$, $|-1,-1/2\rangle$ and $|-2,1/2\rangle$,
$|-1,1/2\rangle$ and $|-2,3/2\rangle$, $|-1,3/2\rangle$ and
$|-2,5/2\rangle$. The bright state with the most important component
$|-1,5/2\rangle$ and the dark state with $|-2,-5/2\rangle$ have no
anticrossings.

\begin{figure}
\includegraphics{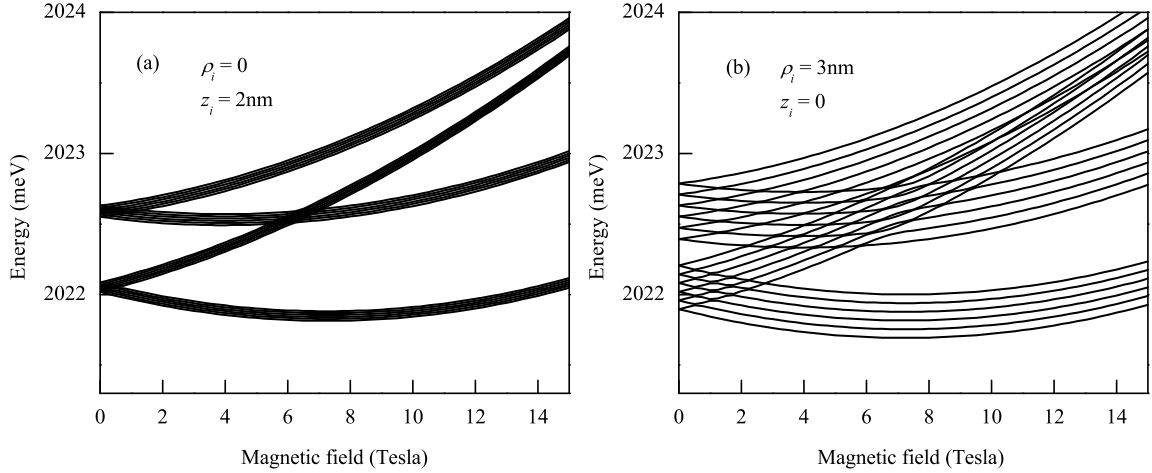}
\caption{\label{fig:OffCenter} Magnetic field dependence of the
electron and hole total energy levels taking into account the sp-d
interaction with an off-center magnetic impurity. (a) Impurity
shifted from the center in grough direction ($\rho_i=0$, $z_i=2$nm).
(b) Impurity shifted from the center in plane direction
($\rho_i=3$nm, $z_i=0$)}
\end{figure}

The two energy states in the inset of figure~\ref{fig:Energy} are
superpositions of $|-1,-5/2 \rangle$ and $|-2,-3/2\rangle$. At low
magnetic fields the main component of the higher level is
$|-1,-5/2\rangle$ and the weight of $|-2,-3/2\rangle$ is much
smaller. Near the anticrossing point the weights of both components
are equal, and for high magnetic fields the main component is
$|-2,-3/2\rangle$. The opposite picture can be seen for the lower
level. This change of the most important component with an increase
of the magnetic field will manifest itself in the optical spectrum
of the system, which we discuss below.

In Figure~\ref{fig:OffCenter} the electron-hole energy levels as a
function of magnetic field are presented for the case of an off-center
impurity. In (a), the impurity is shifted from the center of the dot
in growth direction by 2 nm, and in (b) impurity is shifted in the plane
by 3 nm. In both cases, the energy splitting due to the s,p-d spin
interactions become smaller. This is because the strength of
short range spin interaction depends on the probability to find the
electron (the hole) at the point of the impurity. Since the ground
state electron (hole) is mostly located in the central part of the dot,
if we move the impurity out of the central part, all the
effects described above will become weaker.

In order to calculate the optical transition probabilities, let us
note that the initial state of the system is that of the magnetic
impurity spin with the valence band states fully occupied and the
conduction band states empty. Let us also assume that the impurity
states are pure states $|i\rangle=|S_z\rangle$. Recently, there were
several experimental reports where quantum dots with a single
magnetic impurity in a pure spin state were prepared even in a zero
magnetic field \cite{Reiter, Besombes2009}. The final states are the
eigenstates of the Hamiltonian (\ref{Ham}) presented in
(\ref{Bazis}) $|f\rangle=|\Psi_{J_z} \rangle$. In the electric
dipole approximation the relative oscillator strengths for all
possible optical transitions are proportional to
\begin{equation}\label{Osc}
P(m)\sim\left|\langle\Psi_{J_z}|m,S_z\rangle \right|^2.
\end{equation}
Here the values of $m=1,0,-1$ characterize the polarization of the
light as $\sigma^+$, $\pi$ and $\sigma^-$ respectively
\cite{Bhattacharjee2003}. It should also be mentioned that the
impurity spin state remains unchanged during the optical
transitions.

\begin{figure}
\includegraphics{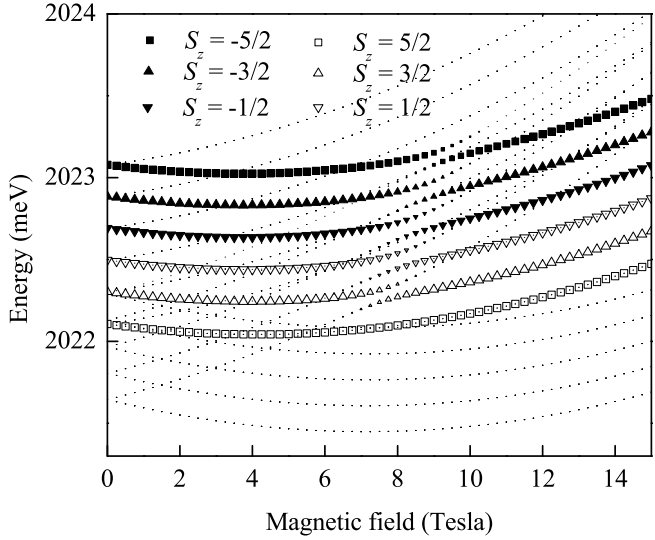}
\caption{\label{fig:optical} Optical transition energies versus the
magnetic field. The shape of the points indicates the impurity spin
of initial state, while the size of the points illustrate the
probability of the transition.}
\end{figure}

In figure~\ref{fig:optical} all possible optical transition energies
of the system are presented as a function of the magnetic field for
$\sigma^-$ polarization of the light and for six different initial
states of the magnetic impurity. The shapes of the points indicate
the initial spin of the impurity and the sizes indicate the
probability of transition. In the case of the initial state with
$S_z=5/2$ the optical transition is possible only to the final
states which have projection to bright exciton state
$|-1,5/2\rangle$ (white squares in figure~\ref{fig:optical}). We
only have one state with the most important component
$|-1,5/2\rangle$ for all values of the magnetic field. We thus see
only one line with high transition probability. For the initial
state $S_z=-5/2$ the transitions are possible only to the states
with $|-1,-5/2\rangle$ as the main component (black squares). Here,
at low magnetic fields we again have only one transition energy
line. But beyond the field of 9 Tesla that line disappears and a new
optical mode appears. Similar behavior can also be seen for other
impurity spin states. This effect is the direct signature of level
anticrossings described above. Near the anticrossing, the weight of
$|-1,-5/2\rangle$ decreases and the dark state $|-2,-3/2\rangle$
becomes the most important component of the wave function. That is
why the bright state changes to the dark state, and for the state
with $|-2,-3/2\rangle$ as the main component, we see an opposite
behavior. After the anticrossing it becomes bright with
$|-1,-5/2\rangle$ as the main component. Similar results obtained
from the observed PL spectra of QDs with single magnetic impurity
were reported by Besombes et al. \cite{Besombes2005}. The six lines
with higher transition probability, presented in
figure~\ref{fig:optical} correspond to the six emission lines for
the $\sigma^-$ polarization presented in Fig.2 (a) of ref.
\cite{Besombes2005}. In both cases we have six lines in the energy
range of 1meV. Further, we show the presence of five level
anticrossings visible in the experiment and describe the underlying
physics. Therefore our theory describes the experimental results of
\cite{Besombes2005} taking into account all the spin interactions
between the electron, the hole and the magnetic impurity and
including the heavy hole - light hole band mixing effects.

Recently, investigations of this type of exciton transitions in
magnetic quantum dots became very attractive because they can be
used as a tool to tune the spin of the magnetic impurity. In a
recent work, Reiter et al. \cite{Reiter} presented an interesting
mechanism for all-optical spin manipulation of a single Mn atom in
the CdTe quantum dot. They prepared a dot with a Mn atom in the pure
state $-5/2$. Such dots can be prepared at low temperatures in an
external magnetic field \cite{Reiter}, or by optical pumping
mechanism presented in \cite{Besombes2009}. Using the laser pulse
with $\sigma^-$ polarization they then created an exciton in the
state $|-1,-5/2\rangle$. That state is coupled with the state
$|-2,-3/2\rangle$ and hence the exciton will perform small Rabi
oscillations between those two states. This Rabi oscillations alone
can not create a spin flip of the Mn atom, but these authors have
shown that the Mn spin can be efficiently controlled by exciting the
dot with a series of laser pulses applied at time intervals given by
half the Rabi periods. After a large number of these pulses, the
impurity spin flips from $S_z=-5/2$ to $S_z=-3/2$. The authors then
did the same experiment at a magnetic field of 9 Tesla (near the
level anticrossing point), and found that only few pulses are
sufficient to bring the exciton into the dark state with spin
$-3/2$. In the same way all the remaining spin states of the Mn atom
can be reached.

In the light of our results presented above, we propose an
alternative route to the mechanism proposed by Reiter et al.:
Magneto-optical mechanism to control the spin of the magnetic
impurity in a QD. Let us consider a QD with a single magnetic
impurity in the pure state $S_z=-5/2$, in an external magnetic field
below the anticrossing point. We can excite the system using the
laser pulse with $\sigma^-$ polarization to create an exciton in the
bright state $|-1,-5/2\rangle$. We then increase the magnetic field
above the anticrossing point. The exciton, by passing through the
anticrossing point will go to the dark state and the main component
of its wave function will be $|-2,-3/2\rangle$ (see
figure~\ref{fig:optical}, black squares) and the spin of the Mn atom
will flip to $-3/2$. Alternatively, we can excite the system in a
magnetic field above anticrossing point, and then decrease the field
to achieve a similar result. Likewise we can go through all the
remaining values of the spin of the Mn atom, as in \cite{Reiter}. We
believe that our scheme will avoid the complex process involving
fine-tuned laser pulses to change the occupation of spin states,
simply by changing the strength of the applied magnetic field. In
fact, if we focus in the region of level anticrossing, the spin flip
actually takes place in a magnetic field range of less than a Tesla.
At the same time the exciton lifetime at low temperatures in self
assembled QDs is of the order of a microsecond \cite{Donega}.
Therefor to generate a spin flip, a change of field of $<1T/\mu$s
would be required. This is currently a technologically challenging
task, but is perhaps achievable in the foreseeable future. Finally,
properties of quantum dots containing a Mn impurity studied here
have the potential for applications in information storage and
read-out. An excellent review on this topic can be found in
\cite{LeGall2010}.

\ack
The work was supported by the Canada Research Chairs Program
and the NSERC Discovery Grant.

\end{document}